\documentclass[%
reprint,
superscriptaddress,
frontmatterverbose,
showpacs,preprintnumbers,
 amsmath,amssymb,
 aps,
pra,
]{revtex4-1}
\usepackage{subfigure}
\usepackage{xcolor}
\usepackage{graphicx}
\usepackage{dcolumn}
\usepackage{bm}
\usepackage{amsmath}
\usepackage{float}
\usepackage[utf8]{inputenc}

\begin{document}

\title{Supersolid behavior in one-dimensional self-trapped Bose-Einstein condensate}

\author{Mithilesh K. Parit}
\email{mithilesh.parit@gmail.com}
\affiliation{Department of Physical Sciences, Indian Institute of Science Education and Research Kolkata, Mohanpur-741246, West Bengal, India}

\author{Gargi Tyagi}
\email{gargityagi150@gmail.com}
\affiliation{Institute of Applied Sciences, Amity University, Sector-125, Noida-201303, Uttar Pradesh, India}

\author{Dheerendra Singh}
\email{ds17ip010@iiserkol.ac.in}
\affiliation{Department of Physical Sciences, Indian Institute of Science Education and Research Kolkata, Mohanpur-741246, West Bengal, India}

\author{Prasanta K. Panigrahi}%
\email{pprasanta@iiserkol.ac.in}
\affiliation{Department of Physical Sciences, Indian Institute of Science Education and Research Kolkata, Mohanpur-741246, West Bengal, India}
\date{\today}
\begin{abstract}

Supersolid is an exotic state of matter, showing crystalline order with a superfluid background, observed recently in dipolar Bose-Einstein condensate (BEC) in a trap. Here, we present exact solutions of the desired Bloch form in the self-trapped free-floating quantum fluid. Our general solutions of the amended nonlinear Schr{\"o}dinger equation,  governing the mean-field and beyond mean-field dynamics, are obtained through a M{\"o}bius transform, connecting a wide class of supersolid solutions to the ubiquitous cnoidal waves. The solutions yield the constant condensate, supersolid behavior and the self-trapped droplet in different parameter domains. The lowest residual condensate is found to be exactly one-third of the constant background. 

\end{abstract}

\maketitle

\section{Introduction} 
Possible existence of the supersolid phase of matter has been conjectured quite some time back \cite{gross, lifs}, with the liquid Helium being the one extensively investigated for its possible realization \cite{kim}. A supersolid is a state of matter that manifests a crystalline order with the presence of a superfluid component, the latter displaying zero viscosity and flow without resistance. Observation in liquid Helium being inconclusive, alternate avenues to its potential observation emerged in the area of cold atomic systems \cite{6r3,Prok}.

Very recently, the transient supersolid phase has been identified in dipolar condensates in different experiments \cite{chomaz1, guo1, 6r2, 6r3}. Three distinct regimes of ground state phase diagram$-$ a regular Bose-Einstein condensate (BEC), coherent and incoherent arrays of quantum droplets have been revealed. B\"ottcher et al. \cite{6r3} and Tanzi et al. \cite{6r2} achieved the supersolid phase in strongly dipolar BECs, arising due to competitive repulsive short-range and attractive long-range dipolar interactions. This has been theoretically well supported by Roccuzzo and Ancilotto \cite{roccuzzo}. In both the experiments, BEC from the trap is released to expand, with the interference of matter waves producing crystalline structure, immersed in a superfluid BEC. The properties of the dipolar supersolid are found to be well described by the extended Gross-Pitaevskii equation, describing the mean-field dynamics with the presence of an additional quartic term \cite{6r3}. Chomaz et al. identified the supersolid properties in the dipolar quantum gases of ${}^{166}{Er}$ and ${}^{164}{Dy}$ \cite{{chomaz1}}.

Guo et al. \cite{guo1} observed two low energy Goldstone modes, revealing the phase rigidity, confirming that array of quantum droplets are in the supersolid phase.  The formation of the supersolid phase is identified through careful observation of these two different Goldstone modes, corresponding to the distinct sound modes of the residual condensed phase and the supersolid vibrations. However, the presence of the trap led to distortion of Goldstone modes. As mentioned earlier, in the case of the other two experiments, the BEC from the trap was released for the matter-waves to interfere and form transient supersolid phase. Hence, it is worth investigating the possibility of realizing supersolid phase without a trap in ultracold scenario.

Recently, Petrov \cite{1r5}, has identified a parametric domain in a two component BEC, where self-trapped quantum droplets form \cite{1r5_ptr}. They have been observed in the free-floating condition \cite{tarruellt, 2r2_ptr,1r7} in dipolar Bose gas and magnetic quantum liquid \cite{3r4}, making them an ideal ground to explore the possibility of the self-trapped quantum liquid in the absence of a trap. Balance of the mean-field (MF) energy and Lee-Huang-Yang (LHY) beyond mean-field \cite{lee} quantum fluctuations (QFs) leads  to these self-trapped quantum liquid droplets. Chomaz et al., demonstrated that LHY stabilization is a general feature of strongly dipolar gases and also examined the role of QFs determining the system properties, particularly its collective mode and expansion dynamics \cite{chomaz2}. Various aspects of the quantum droplets have been explored: Bose-Bose droplets in dimensional crossover regime \cite{1r12}, Bose-Fermi droplets of attractive degenerate bosons and spin polarized fermions \cite{1r13}, have been investigated, some of which have found experimental verification. Bright solitons to droplet transitions have also been demonstrated \cite{1r14}. The case of one dimension is distinct from the two and three dimensions \cite{1r5, 1r25}, as mean-field repulsion is required to balance the quantum pressure, unlike that of attraction in higher dimensions.

The cigar shaped self-trapped quantum droplets have been modelled by the amended nonlinear Schr{\"o}dinger equation (NLSE), which has a repulsive mean-field nonlinearity like regular BEC and a quadratic nonlinearity arising from QFs. It has an exact flat-top large droplet solutions \cite{1r5_ptr} and may have multiple droplet solutions indicated from variational approach \cite{1r5_boris}. 

Here, we have identified exact Bloch wave solutions of the amended nonlinear Schr{\"o}dinger equation, necessarily accompanied by a non-vanishing condensate density, showing clearly the supersolid phase of quantum droplet system. We employ M{\"o}bius transform \cite{panigrahi2, vivek1} to connect a wide class of desired solutions with the cnoidal functions, representing nonlinear periodic density waves, which manifest in diverse physical systems. Our procedure yields a unified picture of the uniform condensate, self-trapped and periodic supersolid phases in different parameter domains.

The paper is organized as follows. In Sec. II, theory of quantum droplets is briefly described, leading to the amended NLSE, governing its dynamics in one dimension. In Sec. III, we present supersolid solutions; a Bloch wave, necessarily possesing a constant background. Their properties in various configurations are highlighted. Finally, we conclude with the summary of results and future directions for investigation.

\section{Dynamics of quantum droplets in one dimension}

Quantum droplets have been shown to occur in the binary BECs, with repulsive intra- ($g_{\uparrow \uparrow},~g_{\downarrow \downarrow}$) and attractive inter-component interaction ($g_{\uparrow \downarrow}$), in the vicinity of the MF collapse instability point. Specifically, they appear in the regime, $0<\delta g = g_{\uparrow \downarrow}+\sqrt{g_{\uparrow \uparrow}~g_{\downarrow \downarrow}} \ll \sqrt{g_{\uparrow \uparrow}~g_{\downarrow \downarrow}}$, with attractive inter-component interaction $g_{\uparrow \downarrow}<0$ and repulsive intra-component average interaction $g=\sqrt{g_{\uparrow \uparrow}~g_{\downarrow \downarrow}}>0$. The energy density of such homogeneous mixture has been obtained \cite{1r5_ptr},

\begin{multline}
    \mathcal{E}=\frac{(g_{\uparrow \uparrow}^{1/2} n_{\uparrow}-g_{\downarrow \downarrow}^{1/2} n_{\downarrow})^2}{2}\\
    + \frac{\sqrt{g_{\uparrow \uparrow}~g_{\downarrow \downarrow}}~ (g_{\uparrow \downarrow}+\sqrt{g_{\uparrow \uparrow}~g_{\downarrow \downarrow}})~ (g_{\uparrow \uparrow}^{1/2} n_{\uparrow}+g_{\downarrow \downarrow}^{1/2} n_{\downarrow})^2 }{(g_{\uparrow \uparrow} + g_{\downarrow \downarrow})^2}\\
    -\frac{2\sqrt{M}}{3\pi \hbar} (g_{\uparrow \uparrow} n_{\uparrow}+g_{\downarrow \downarrow} n_{\downarrow})^{3/2},
    \label{energy1}
\end{multline}
where $n_{\uparrow}$ and $n_{\downarrow}$ are densities of the two components, related by $n=n_{\uparrow} = n_{\downarrow}\sqrt{g_{\uparrow \uparrow}/g_{\downarrow \downarrow}}$. The first two terms in Eq. (\ref{energy1}) are the MF contribution and the last term represents the LHY-BMF correction. With the assumption, $g=g_{\uparrow \uparrow} \sim g_{\downarrow \downarrow}$, we have $n = n_{\uparrow} = n_{\downarrow}$ and $0<\delta g \ll g_{\uparrow \uparrow} \sim g_{\downarrow \downarrow}$. The energy density reduces to,

\begin{equation}
   \mathcal{E} = \frac{\delta g n^2}{2}-\frac{2\sqrt{M}}{3\pi \hbar} \left(gn\right)^{3/2},
 \end{equation}
having the equilibrium density $n_0 = 8g^3/(9 \pi^2 \delta g^2)$ and chemical potential $\mu_0 = -\delta g n_0/2$. The validity of $\mathcal{E}$ has been justified by diffusion Monte-Carlo simulation \cite{1r5_ptr}.

The modified Gross-Pitaevskii equation, with cubic MF and quadratic BMF nonlinearty, has the form,


\begin{equation}
    i\hbar~ \psi_t= -\frac{\hbar^2}{2M}~ \psi_{xx} +\delta g~ \left|\psi\right|^2\psi- \frac{\sqrt{2M}}{\pi \hbar}g^{3/2} \left|\psi\right|\psi. 
    \label{eqn:1}
\end{equation}
It has constant solution: $\psi_{const} = \frac{\sqrt{2M}}{\pi \hbar}\frac{g^{3/2}}{\delta g}$, arising from the balance of MF and BMF contributions.  The plane wave solution, $\psi=\sqrt{P}\exp{(ikx-i \frac{\mu}{\hbar} t)}$, which is prone to modulational instability (MI), with 

\begin{equation}
    \sqrt{P}_{\pm}= \frac{\sqrt{2M} g^{3/2}}{2 \pi \hbar~ \delta g} \pm \sqrt{\frac{M g^3}{2 \pi^2 \hbar^2~ \delta g^2} + \frac{\bar{\mu}}{\delta g} },
\end{equation}
where $\bar{\mu}=\left(\mu - \frac{\hbar^2 k^2}{2M} \right)$. The steady-state solution is perturbed by small amplitude and the resulting MI has been investigated in \cite{mi1, mi2}. MI manifests at wave number $k_P$, where $\Omega$ becomes complex
\begin{equation}
   \Omega=\pm~ \frac{1}{2M}~\sqrt{k_P^2 \left(\hbar^2 k_P^2-4M\left(\frac{\sqrt{2M}}{\pi \hbar}g^{3/2} \frac{\sqrt{P}}{2} - \delta g P\right)\right)},
   \label{disp1}
\end{equation}
with $k_P$ is the wavenumber of the small perturbation. MI occurs for $g>\left(\frac{2\pi^2 \hbar^2 P}{M} \delta g^2\right)^{1/3}$ and

\begin{equation}
    \left|k_P\right|<2 \sqrt{\left(\frac{\sqrt{2M}}{\pi \hbar}g^{3/2} \frac{\sqrt{P}}{2} - \delta g P\right)} \equiv k_0.
\end{equation}
MI is maximum at $k_P=k_0/\sqrt{2}$ and ${k_0}$ is maximum at $\mu=-\frac{3M}{8~\pi^2 \hbar^2}\frac{g^3}{\delta g}$. The regime in the parameter domain, where MI occurs is conducive for inhomogeneous solutions. In the following section, we investigate the possible inhomogeneous solutions for the amended NLSE, explicitly showing the presence of a supersolid phase. 


\section{Supersolid phase}   
We consider propagating Bloch function type solutions \cite{kittel}:
\begin{equation}
\psi=\psi_0\left(\frac{x-vt}{\xi}, m\right)\exp\left(ikx-i\frac{\mu}{\hbar}t\right).
\label{ans:mu}
\end{equation}
Here $v=\frac{\hbar k}{M}$ and $\psi_0$ is a periodic function with m being the modulus parameter:
\begin{equation}	
-\frac{\hbar^2}{2M}~ \partial^2_x \psi_0 + \delta g \psi_0^3- \frac{\sqrt{2M}}{\pi\hbar} g^{3/2} \psi_0^2-\bar \mu \psi_0=0.
\end{equation}
Henceforth, we take $\psi_0$ to be a real periodic function without loss of generality, taking advantage of the global $U(1)$ symmetry of the system. For finding general solutions, appropriate M\"obius transformations are employed, to connect the solution space to the ubiquitous cnoidal waves \cite{hancock}, satisfying: $f"\pm af\pm \lambda f^3 = 0$ \cite{panigrahi2}. The general solution has the form,

\begin{equation}
\psi_0(x,t)=\frac{A + Bf^\delta(\frac{x-vt}{\xi} )}{1 + Df^\delta(\frac{x-vt}{\xi} )}
\label{sol_f1}
\end{equation}
Balancing of nonlinearity with dispersion leads to $\delta=1$ and $\delta=2$ as possible choices. Here, we consider $\delta=1$ as it has richer structure, which allows for diverse boundary conditions to be satisfied. We now start with the periodic solution, 

\begin{equation}
    \psi_0 \left(x, t\right)=A+B~\text{sn}\left(\frac{x-vt}{\xi},m\right),
    \label{ss_sn}
\end{equation}
and find that it is necessarily accompanied by positive background $A = \frac{\sqrt{2M}}{3\pi \hbar}\frac{g^{3/2}}{\delta g}$, having one-third the value of the uniform background. It is worth emphasizing that the supersolid solutions does not smoothly trace it to the constant condensate in the limit $m\to0$ when $B=0$. Its amplitude,  $B=\pm\sqrt{\frac{2m}{m+1}}A$, never exceeds that of the background in the allowed energy range of the modulus parameter $0< m <1$. The healing length is obtained as
\begin{equation}
   \xi=\frac{3\hbar^2 \pi}{2M}\sqrt{\frac{(m+1)~ \delta g}{g^3}},
    \label{eqn xi}
\end{equation}
It is evident that, the existence of the Bloch type solutions with a superfluid background crucially depends upon the cubic non-linearity, BMF correction and dispersion.
For a moving supersolid, $ \frac{\hbar^2 k^2}{2M}-\mu = \frac{4M}{9 \pi^2 \hbar^2}\frac{g^3}{\delta g}$, revealing that the chemical potential is bounded below, $\mu_{min}=\mu_0=-\frac{4M}{9 \pi^2 \hbar^2}\frac{g^3}{\delta g}<0$, identical to that of the self-trapped droplet \cite{ 1r5_ptr}.
Here, $\psi_{min/max}=A\pm B=A\left(1\pm\sqrt{\frac{2m}{m+1}}\right)>0$, showing clearly that for the quantum supersolid immersed in a residual BEC \cite{6r2, 6r3, roccuzzo}. The density of the residual BEC  $n_{res}=A^2\left(1-\sqrt{\frac{2m}{m+1}}\right)^2>0$. This diffused matter-wave density rules out the scenario of one atom per site thereby overtaking the Penrose and Onsagar criterion \cite{PenOn, chester, Dy1994}.

 \begin{figure}[ht]
\includegraphics[width=9.50 cm,height=9.0 cm]{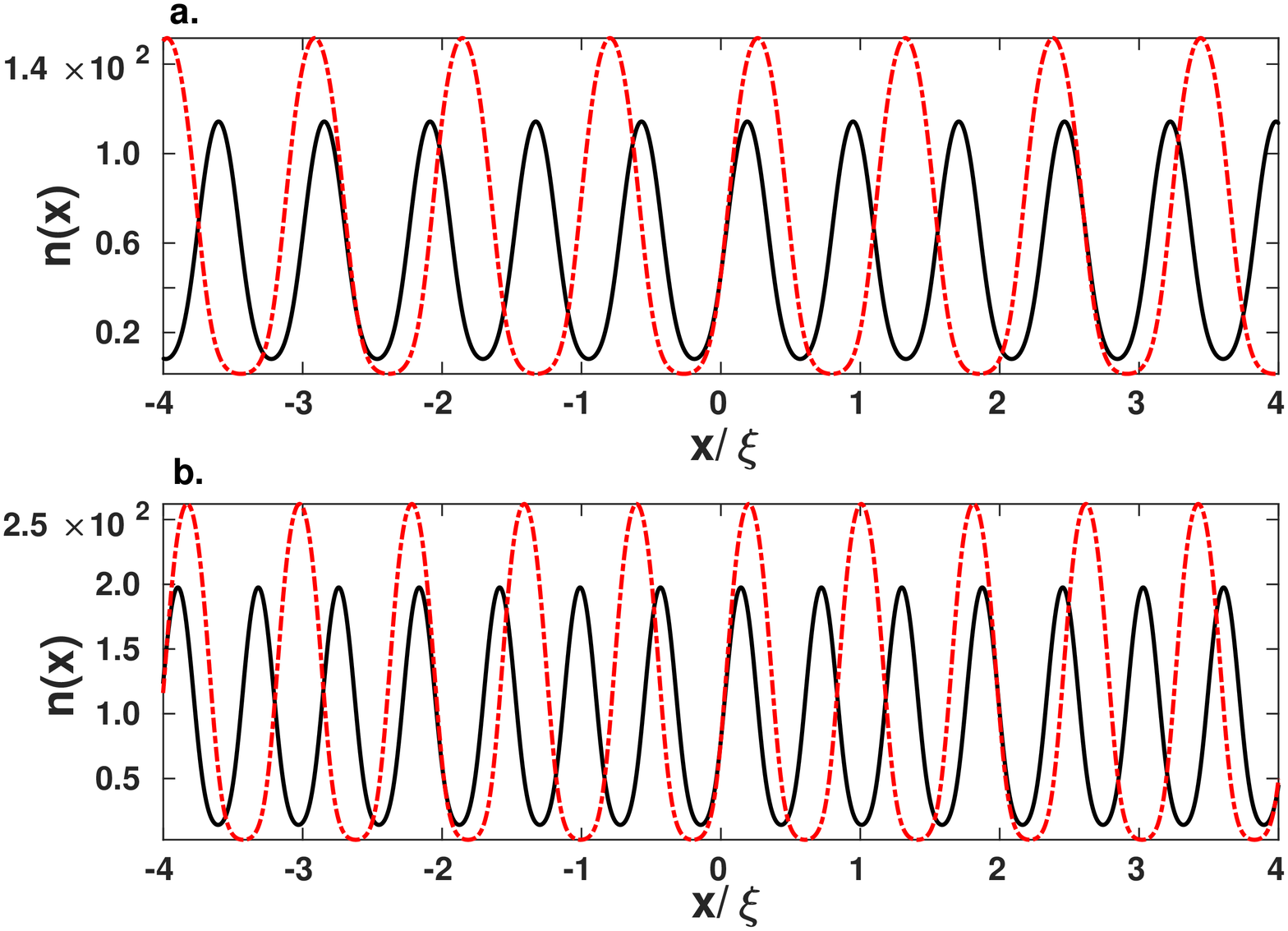}\\
\includegraphics[width=9.50 cm,height=3.0 cm]{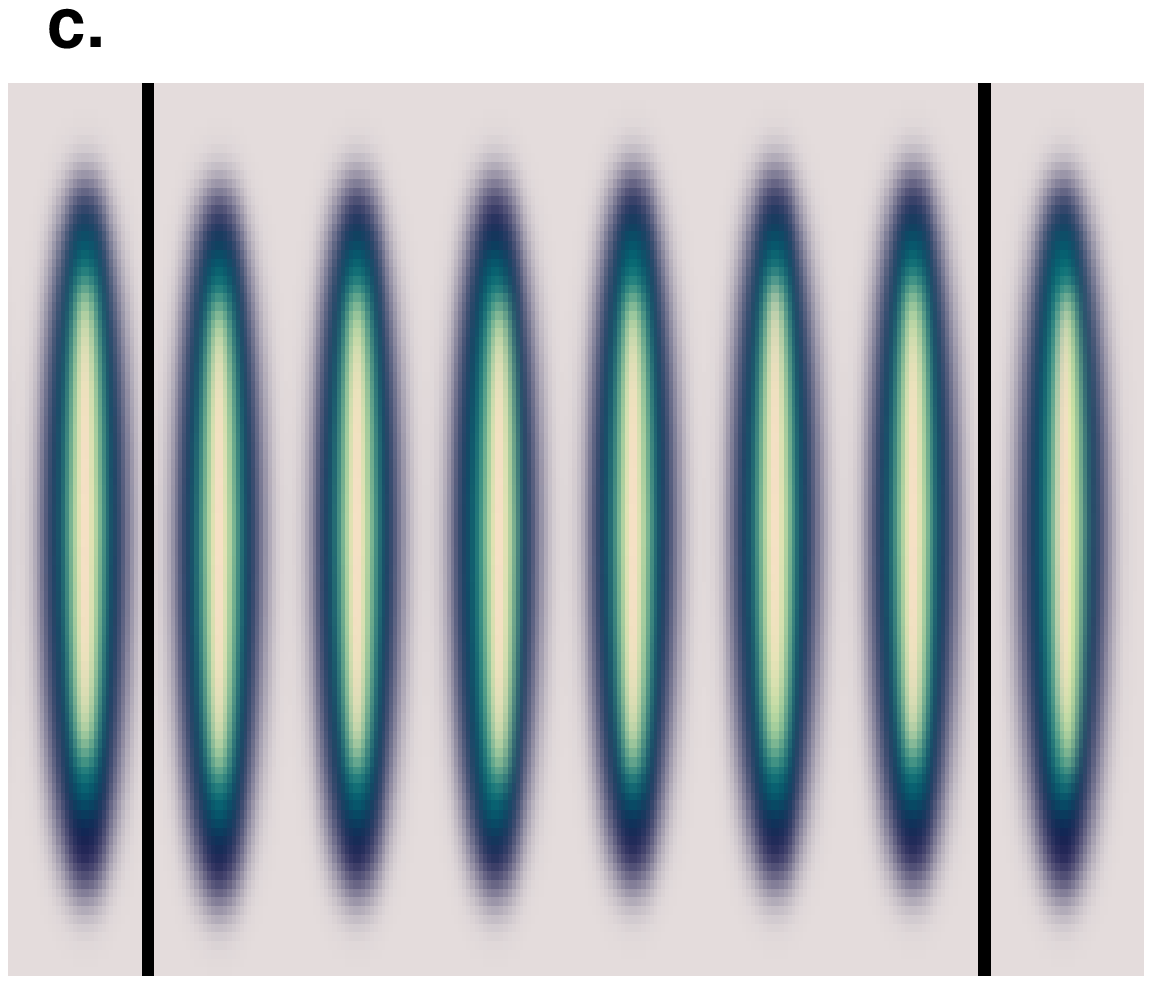}
\centering
\caption{(Color online) The density of supersolid phase as a function of scaled position is depicted for different values of intra-component interaction ($g$) and  modulus parameter, with $\delta g=0.1$ (black: $m=0.2$ and red: $m=0.5$), for both the panels. The value of intra-component interaction, $g=10$ for panel (a) and $g=12$ for panel (b). Panel (c) shows the pictorial representation of several quantum droplets immersed in a residual BEC, with boundaries from the both sides (solid vertical lines), confined in a box of finite length.}
\label{ssphase1}
\end{figure}

The density of the supersolid phase is shown in Fig. \ref{ssphase1}, as a function of scaled position variable $x/\xi$, in the comoving frame. It is evident that inter-droplet spacing (peak-to-peak distance) and number of quantum droplets are inter-related. With increasing repulsive intra-component interaction $g$, number of droplets increases and their size decreases for a fixed value of $x/\xi$. Further, keeping all the physical parameters constant, increase in the modulus parameter increases the inter-droplet spacing. 

The second exact solution is of the sinusoidal form: 

\begin{equation}
    \psi_0=\frac{A}{1+D~ sin\left(\frac{x-vt}{\xi}\right)},
    \label{eqn:sol2}
\end{equation}
with the amplitudes, $A=\frac{3 \pi \hbar \bar{\mu} }{\sqrt{2M}g^{3/2}}$ and $D=\sqrt{1-\frac{9 \pi^2 \hbar^2}{4M}\frac{\delta g}{g^3} \bar{\mu}}$. The healing length $\xi=\sqrt{-\frac{\hbar^2}{2M \bar\mu}}$ and $\mu_{min}=-\frac{4M}{9 \pi^2 \hbar^2}\frac{g^3}{\delta g}$. Hence, the minimum chemical potential is again identical to that of the self-trapped droplet \cite{ 1r5_ptr}.

We now proceed to study the general periodic solution of the amended NLSE,
\begin{equation}
\psi_0=\frac{A+B~ sn\left(\frac{x-vt}{\xi}, m\right)}{1+D~ sn\left(\frac{x-vt}{\xi}, m\right)},
\label{sol_sn1}
\end{equation}
A straightforward but tedious calculation yields, $D=\pm \sqrt{2m/(m+1)}$, $B=DX_{\pm}$, with $X_{\pm} = \left(g_2\pm \sqrt{g_2^2-4 g_1 \mu}\right)/2 g_1$. The value of $D$, $0<|D|<1$, shows the non-singular nature of the general solutions. The possible solutions of $A$ and $\xi$ ($\frac{1}{\xi^2} = \frac{M L}{\hbar^2}$) are listed below:

\begin{subequations}
\begin{align}
    \label{vA4}
    A_{\pm}^1&=\frac{g_2 X_+ \pm 6 \sqrt{\left(g_2^2/4 g_1 - \bar{\mu} \right) \left(g_2 X_+ - g_2^2/2 g_1 - \bar{\mu} \right)}}{g_2 + 3 g_1 \sqrt{g_2^2 - 4 g_1 \bar{\mu} }},\\[5pt]
    \label{vA5}
    A_{\pm}^2&=\frac{g_2 X_- \pm 6 \sqrt{\left(g_2^2/4 g_1 - \bar{\mu} \right) \left(g_2 X_-  - \bar{\mu} \right)}}{g_2 - 3 g_1 \sqrt{g_2^2-4 g_1 \bar{\mu} }},\\[5pt]
    \label{vl3}
    L_{\pm}&=-\frac{g_2 X_+/2 - \bar{\mu} \pm 3 \sqrt{\left(g_2^2/4 g_1 - \bar{\mu} \right) \left(g_2 X_+ - \bar{\mu} \right)}}{(m+1)},\\[5pt]
    \label{vl4}
    L_{\pm}&=\frac{g_2 X_+/2 - \bar{\mu} \pm 3 \sqrt{\left(g_2^2/4 g_1 - \bar{\mu} \right) \left(g_2 X_- - \bar{\mu} \right)}}{(m+1)}.
\end{align}
\end{subequations}
The chemical potential: (i) $\mu>-\frac{g_2^2}{4g_1}$ and (ii) $-\frac{0.414~g_2^2}{g_1}<\mu<0$. The minimum chemical potential is less than that of obtained for the self-trapped supersolids [Eq. (\ref{ss_sn}) and (\ref{eqn:sol2})] or quantum droplets \cite{1r5_ptr}.

It is worth mentioning that, for the case of cigar-shaped BEC, the Gross-Pitaevskii (GP) equation does not yield  a constant background with  periodic modulation. The presence of both mean-field and BMF energies are crucial for the existence of these solutions here. The soliton trains observed in the attractive BEC are composed of BEC droplets in the absence of a superfluid background \cite{Partridge}. For the case of NLSE with a phase locked source, periodic solutions with constant background are permissible \cite{panigrahi2}. In optical fibers analogous periodic solutions in the temporal domain have been experimentally observed as frequency combs \citep{Gorodetsky}. Recently, NLSE with cubic-quadratic nonlinearity and phase locked with a source has been investigated, wherein the parity even and kink type solutions have been identified  \cite{Kumar}.  The cubic-quadratic equation with a phase-locked source have also been investigated through the fractional linear transformations, for attractive Kerr nonlinearity and repulsive BMF correction \cite{belicd}. The behavior of solitons under nonlinearity management has been explored \cite{malomed2}. In the ensuing sections, the dispersion relation for supersolid is investigated, distinctly showing that increase of repulsive inter-component interaction decreases the energy.

\section{Energy and momentum of the supersolid phase}
We now investigate the dispersion relation of the supersolid phase. The energy and momentum density \cite{pethick, 1r27} can be computed for a supersolid in a box of finite length. The expressions for the momentum and energy are provided in the supplementary information.

\begin{figure}[H]
\includegraphics[width=8.5 cm,height=8.0 cm]{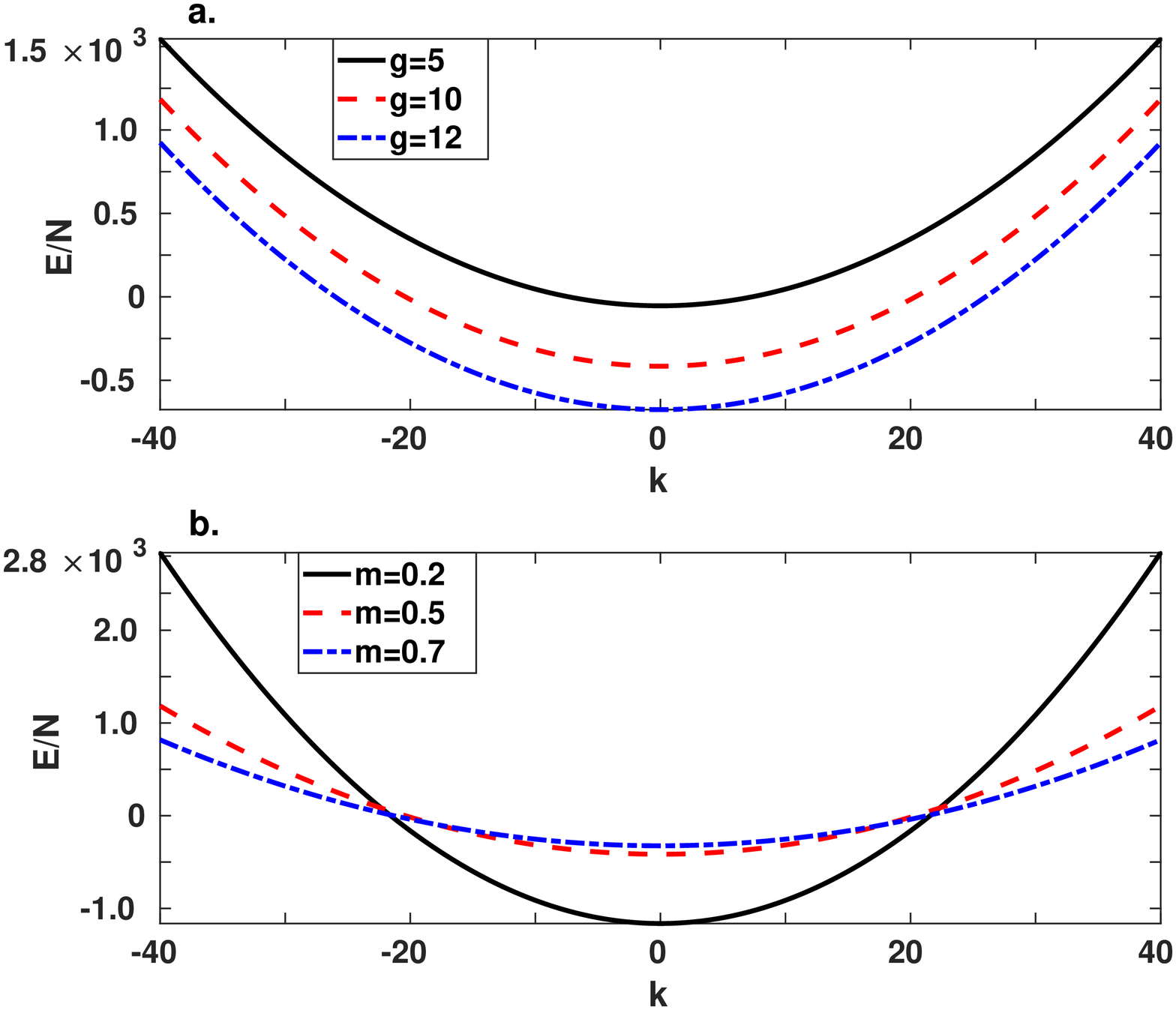}
\centering
\caption{(Color online) The energy per atom of the supersolid as a function of $k$ is depicted for different values of repulsive intra-component interaction and modulus parameter, with $m=0.5$. It shows that increasing repulsive interaction while keeping the competitive intra- and inter-component interaction constant, $\delta g=0.1$, decreases the total energy per atom. The value of intra-component interaction is $g=10$ for panel (b). Increasing the value of modulus parameter leads to broader dispersion, due to periodicity of the solution.}
\label{fig:disp1}
\end{figure}

The dispersion relation for supersolid phase is demonstrated in Fig. (\ref{fig:disp1}), for different repulsive inter-component interaction in panel $a$. It exhibits the decrease in energy per atom with increasing $g$. In panel $b$, the dispersion relation is shown for different moduli parameter, showing comparative flattening of the energy per atom for greater moduli. This happens due to formation of front-like droplet, with infinite period when modulus parameter approaches unity. 


\section{Conclusion}
In conclusion, we have obtained the exact supersolid solutions in one-dimensional self-trapped BEC. The supersolid phase occurs in a free-floating condition, unlike the case of dipolar BEC, wherein the condensate with appropriately tuned interactions needs to be released, and the interference of matter-wave results in supersolid formation with superfluid background in the exact periodic solution. In the present case, exact Bloch waves, quantum droplets appear in a periodic array with inter-spacing between two droplets constant, immersed in a residual BEC background, that establishes inter-droplet coherence \cite{6r3, 6r4}. The chemical potential for two solutions is identical and for the general solution it is less, which is more favourable.

\section{Acknowledgements}
We thank Prof.  R. B. MacKenzie for fruitful discussion. M. K. Parit acknowledges the financial support from the \textit{REDX}, Center for Artificial Intelligence, IISER Kolkata, funded by \textit{Silicon Valley Community Foundation} through grant number 2018-191175 (5618). G. Tyagi is thankful to IISER Kolkata for hospitality.


\pagebreak

\onecolumngrid

\section*{Supplementary Information: Supersolid behavior in one-dimensional self-trapped Bose-Einstein condensate}

\subsection*{Energy and momentum of the supersolid phase}
In this section, we have provided the analytical expression for the energy and momentum. The energy and momentum density \cite{pethick, 1r27} can be computed, $E=\int \mathcal{H}~dx$, where $\mathcal{H}$ is the Hamiltonian density
\begin{equation}
    \mathcal{H}=\frac{\hbar^2}{2M}\left|\partial_x \psi \right|^2+\frac{\delta g}{2}~\left|\psi\right|^4 - \frac{2\sqrt{2M}}{3 \pi \hbar}g^{3/2} \left|\psi \right|^3,
    \label{ham_d}
\end{equation}
and momentum $P= \frac{i \hbar}{2} \int \left( \psi \partial_x \psi^* - \psi^* \partial_x \psi \right)dx.$ The analytical expressions for the momentum and energy are presented below for the solution-I, with $A = \frac{\sqrt{2M}}{3\pi \hbar}\frac{g^{3/2}}{\delta g}$, $B=\pm\sqrt{\frac{2m}{m+1}}A$, $\xi=\frac{3\hbar^2 \pi}{2M}\sqrt{\frac{(m+1)~ \delta g}{g^3}}$, and $D_1=\frac{\hbar^2}{2M}$.

\begin{enumerate}

\item \textbf{Number of atoms}\\[10pt]
    The number of atoms $N=\int \left|\psi \right|^2dx$.

\begin{multline}
    \mathbf{N(x)}= \left(A^2 m+B^2\right) (x-t v)+B^2 \xi  E\left(\left.\text{am}\left(\left.\frac{t v-x}{\xi }\right|m\right)\right|m\right)  \\  +2 A B \sqrt{m} \xi  \log \left(\text{dn}\left(\left.\frac{x-t v}{\xi }\right|m\right)-\sqrt{m} \text{cn}\left(\left.\frac{x-t v}{\xi }\right|m\right)\right).
\end{multline}

\item \textbf{Energy density}

\begin{subequations}
\begin{align}
    \label{engy1a}
    \mathbf{E(x)}&=\frac{E_1+E_2+E_3+E_4+E_5\times \lambda_E}{6 m^2 \xi ^2}\\
    \label{engy1b}
    E_1&=2 \left(3 A^2 m^2 \xi ^2 (x-t v) \left(A^2 g_1-A g_2+D_1 k^2\right)+B^4 g_1 (m+2) \xi ^2 (x-t v)\right)\\
    \label{engy1c}
    E_2&=2 \left(B^2 m \left(9 A \xi ^2 (2 A g_1-g_2) (x-t v)+D_1 t v \left(-3 k^2 \xi ^2+2 m+1\right)+D_1 x \left(3 k^2 \xi ^2+m-1\right)\right)\right)\\
    \label{engy1d}
    E_3&=2 B^2 \xi  E\left(\left.\text{am}\left(\left.\frac{x-t v}{\xi }\right|m\right)\right|m\right) \left(\xi ^2 \left(-9 A m (2 A g_1-g_2)-2 B^2 g_1 (m+1)\right)+D_1 m \left(-3 k^2 \xi ^2+m+1\right)\right)\\
    \label{engy1e}
    E_4&=B^2 m \xi  \text{cn}\left(\left.\frac{x-t v}{\xi }\right|m\right) \text{dn}\left(\left.\frac{x-t v}{\xi }\right|m\right) \left(3 B \xi ^2 (4 A g_1-g_2)+2 \left(B^2 g_1 \xi ^2+D_1 m\right) \text{sn}\left(\left.\frac{x-t v}{\xi }\right|m\right)\right)\\
    \label{engy1f}
    E_5&=3 B \sqrt{m} \xi ^3 \left(8 A^3 g_1 m-6 A^2 g_2 m+4 A \left(B^2 g_1 (m+1)+D_1 k^2 m\right)-B^2 g_2 (m+1)\right)\\
    \label{engy1g}
    \lambda_E&=\log \left(\text{dn}\left(\left.\frac{x-t v}{\xi }\right|m\right)-\sqrt{m} \text{cn}\left(\left.\frac{x-t v}{\xi }\right|m\right)\right),
\end{align}
\end{subequations}

\item \textbf{Momentum density}

\begin{multline}
    \mathbf{P(x)}=\frac{\hbar k}{m} \left( \left(A^2 m+B^2\right) (x-t v)+B^2 \xi  E\left(\left.\text{am}\left(\left.\frac{t v-x}{\xi }\right|m\right)\right|m\right) \right. \\ \left. +2 A B \sqrt{m} \xi  \log \left(\text{dn}\left(\left.\frac{x-t v}{\xi }\right|m\right)-\sqrt{m} \text{cn}\left(\left.\frac{x-t v}{\xi }\right|m\right)\right) \right)=\frac{\hbar k}{m}\mathbf{N(x)}.
\end{multline}

\end{enumerate}
 For supersolids confined in a box of finite length $L$ (from $-L/2$ to $L/2$): number, $N=N(L/2)-N(-L/2)$, energy, $E=E(L/2)-E(-L/2)$, and momentum, $P=P(L/2)-P(-L/2)$.



\begin{thebibliography}{99}
\bibitem{gross}{E. P. Gross, Phys. Rev. \textbf{106}, 161-162 (1957).}

\bibitem{lifs}{A. F. Andreev and I. M. Lifshitz, Sov. Phys. JETP \textbf{29}, 1107-1113 (1969).}

\bibitem{kim}{D. Y. Kim and M. H. W. Chan, Phys. Rev. Lett. \textbf{109}, 155301 (2012) and references therein.}

\bibitem{Prok}{M. Boninsegni and N. V. Prokof’ev, Rev. Mod. Phys. \textbf{84}, 759-776 (2012).}

\bibitem{6r3}{F. B\"ottcher, J.-N. Schmidt, M. Wenzel, J. Hertkorn, M. Guo, L. Tim, and P. Tilman,  Phys. Rev. X \textbf{9}, 011051 (2019).}

\bibitem{6r2}{L. Tanzi, E. Lucioni, F. Famà, J. Catani, A. Fioretti, C.  Gabbanini, R. N.  Bisset, L.  Santos, and G. Modugno, Phys. Rev. Lett. \textbf{122}, 130405 (2019).}

\bibitem{chomaz1}{L. Chomaz, D. Petter, P. Ilzh\"ofer, G. Natale, A. Trautmann, C. Politi, G. Durastante, R. M. W. van Bijnen, A. Patscheider, M. Sohmen, M. J. Mark, and F. Ferlaino, \textit{Phys. Rev. X} \textbf{9}, 021012 (2019).}

\bibitem{guo1}{M. Guo, F. B\"ottcher, J. Hertkorn, J.-N. Schmidt, M. Wenzel,
H. P. B\"uchler, T. Langen, and T. Pfau, Nature (London)\textbf{ 574}, pp. 386-389 (2019).}

\bibitem{roccuzzo}{A. M. Roccuzzo and F. Ancilotto, Phys. Rev. A \textbf{99}, 041601(R) (2019).}

\bibitem{1r5}{D. S. Petrov, \textit{Phys. Rev. Lett.} \textbf{115}, 155302 (2015).}

\bibitem{1r5_ptr}{D. S. Petrov and G. E. Astrakharchik, \text{Phys. Rev. Lett.} \textbf{117}, 100401 (2016).}

\bibitem{tarruellt}{C. R. Cabrera, L. Tanzi, J. Sanz, B. Naylor, P. Thomas, P. Cheiney, L. Tarruell, \textit{Science} \textbf{359}, pp. 301-304 (2018).}

\bibitem{1r7}{G. Semeghini, G. Ferioli, L. Masi, C. Mazzinghi, L. Wolswijk, F. Minardi, M. Modugno, G. Modugno, M. Inguscio, and M. Fattori, Phys. Rev. Lett. \textbf{120}, 235301 (2018).}

\bibitem{2r2_ptr}{D. S. Petrov, \textit{Nature} \textbf{14}, pp. 211-212 (2018) and references therein.}

\bibitem{3r4}{M. Schmitt, M. Wenzel, F. B\"ottcher, I. F.-Barbut, T. Pfau, \textit{Nature} \textbf{539}, pp. 259-262 (2016).}

\bibitem{lee}{T. D. Lee, K. Huang and C. N. Yang, \textit{Phys. Rev.} \textbf{106}, 1135 (1957).}

\bibitem{chomaz2}{L. Chomaz, S. Baier, D. Petter, M. J. Mark, F. Wächtler, L. Santos, and F. Ferlaino, Phys. Rev. X \textbf{6}, 041039 (2016).}

\bibitem{1r12}{P. Zin, M. Pylak, T. Wasak, M. Gajda, and Z. Idziaszek, \textit{Phys. Rev. A} \textbf{98}, 051603(R) (2018).}

\bibitem{1r13}{D. Rakshit, T. Karpiuk, M. Brewczyk, M. Gajda, \textit{Sci. Post Phys.} \textbf{6}, 079 (2019).}

\bibitem{1r14}{P. Cheiney, C. R. Cabrera, J. Sanz, B. Naylor, L. Tanzi, and L. Tarruell, \textit{Phys. Rev. Lett.} \textbf{120}, 135301 (2018).}

\bibitem{1r25}{V. V. Vyborova, O. Lychkovskiy, and A. N.  Rubtsov, \textit{Phys. Rev. B} \textbf{98}, 235407 (2018).}

\bibitem{1r5_boris}{G. E. Astrakharchik and B. A. Malomed, \textit{Phys. Rev. A} \textbf{98}, 013631 (2018).}

\bibitem{panigrahi2}{T. S. Raju, C. N. Kumar and P. K. Panigrahi, \textit{ J. Phys. A: Math. Gen.} \textbf{38}, L271–L276 (2005).}

\bibitem{vivek1}{V. M. Vyas, T. S. Raju, C. N. Kumar, and P. K. Panigrahi, J.
Phys. A \textbf{39}, 9151 (2006).}

\bibitem{mi1}{T. Mithun, A. Maluckov, K. Kasamatsu, B. A. Malomed, and A. Khare, \textit{Symmetry} \textbf{12} (1), 174 (2020).}

\bibitem{mi2}{D. Singh, M. K. Parit,  T. S. Raju, and P. K. Panigrahi, Res. gate Prepr., doi:10.13140/RG.2.2.34638.82246 (2019).}

\bibitem{kittel}{C. Kittel,\textit{ Introduction to Solid State Physics}, (Wiley: New York,
1996).}

\bibitem{hancock}{H. Hancock,\textit{ Theory of Maxima and Minima,} (Dover publication, New York, 1958).}

\bibitem{PenOn}{O. Penrose and L. Onssger, \textit{Phys. Rev.} \textbf{104,} 576 (1956).}

\bibitem{chester}{G. V. Chester, \textit{Phys. Rev. A} \textbf{2}, 256 (1970).}

\bibitem{Dy1994}{Y. Pomeau and S. Rica, \textit{Phys. Rev. Lett.} \textbf{72}, 2426 (1994) and references therein.}

\bibitem{Partridge}{U. A. Khawaja, H. T. C. Stoof, R. G. Hulet, K. E. Strecker and G. B. Partridge, \textit{Phys. Rev. Lett.} \textbf{89}, 200404 (2002).}

\bibitem{Gorodetsky}{T. J. Kippenberg, A. L. Gaeta, M. Lipson, M. L. Gorodetsky,\textit{ Science} \textbf{361}, eaan8083  (2018).}

\bibitem{Kumar}{R. Pal, H. Kaur, A. Goyal, and C. N. Kumar, J. of Mod. Optics, \textbf{66} (5), pp.  571-579  (2019).}

\bibitem{belicd}{H. Triki, A. Biswas, S. P. Moshokoab and M. Belicd \textit{Optik} \textbf{128}, 63-70 (2017).}

\bibitem{malomed2}{J. Fujioka1, E. Cortés, R. Pérez-Pascual, R. F. Rodríguez, A. Espinosa1, and B. A. Malomed \textit{AIP} \textbf{21}, 033120 (2011).}

\bibitem{pethick}{C. J. Pethick and H. Smith, \textit{Bose Einstein Condensation in Dilute Gases}, (Cambridge University Press, Cambridge, United Kingdom, 2001).}

\bibitem{1r27}{A. D. Jackson, and  G. M. Kavoulakis \textit{Phys. Rev. Lett.} \textbf{89}, 070403 ( 2002) and references therein.}

\bibitem{6r4}{M. Wenzel, F. B\"ottcher, T. Langen, I. F.-Barbut, and T. Pfau \textit{Phys. Rev. A} \textbf{96}, 053630 (2017).}


\end{thebibliography}
\end{document}